# Detecting Resting-state Neural Connectivity Using Dynamic Network Analysis on Multiband fMRI Data

Jiancheng Zhuang

Dornsife Imaging Center, University of Southern California, Los Angeles, CA, USA

**Abstract**

This paper describes an approach of using dynamic Structural Equation Modeling (SEM) analysis to estimate the connectivity networks from resting-state fMRI data measured by a multiband EPI sequence. Two structural equation models were estimated at each voxel with respect to the sensory-motor network and default-mode network. The resulting connectivity maps indicate that supplementary motor area has significant connections to left/right primary motor areas, and medial prefrontal cortex link significantly with posterior cingulate cortex and inferior parietal lobules. The results imply that high temporal resolution images obtained with multiband fMRI data can provide dynamic and directional information on the neural connectivity.

# Introduction

Determining connectivity between cortical areas from resting-state fMRI data has become an important tool in neuroscience. Historically, the primary focus of fMRI researchers has been on localization of neural activity for resting state or particular tasks, even though their neuroscientific interest is often in the distribution of activity or connectivity across different brain regions. Recently, more attention has been paid to the detection of causal interactions between cortical areas. Conventional correlation analysis used in most connectivity mappings cannot provide causal or directional information between multiple regions in the brain (McIntosh and Gonzalez-Lima, 1994; Büchel and Friston, 1997; Goncalves and Hall, 2003; Penny et al., 2004; Zhuang et al., 2005; Smith et al., 2006).

Previously, we have shown that Structural Equation Modeling (SEM) can be an effective method to ascertain path directions and coefficients from the covariance structure in fMRI data (Zhuang et al., 2005 and 2013). Structural equation modeling (SEM) is a statistical technique that is able to examine causal relationships between multiple variables. The parameters in the SEM are connection strengths or path coefficients between different variables, and reflect the effective connectivity in our neural network model. Each path in the model has direction, and a solvable model must have more than two variables. Parameters are estimated by minimizing the difference between the observed covariances and those implied by a structural or path model. Since SEM solves the whole path model at once, the solution will give the causal direction between multiple regions of interest (ROIs).

While SEM is an effective method to ascertain path directions and coefficients from the covariance structure in fMRI data, in the current literature it is mostly applied using a few fixed models or fixed variables (ROIs in functional neuroimaging). Because these fixed models or ROIs need to be predefined based on existing knowledge, unknown brain areas cannot be explored using these methods. The present work uses SEM in an exploratory analysis to derive, on a voxel-by-voxel basis, the most significant path model and the corresponding path weights, such that we can generate maps corresponding to model statistical indices, especially finding those areas in which the pattern is uncertain in the connectivity analysis.

The usefulness of SEM depends on the dynamic content of the data. The recently developed mutilband slice-accelerated technique provides higher temporal resolution for measuring the dynamic fluctuation of fMRI BOLD signals than conventional EPI sequence (Feinberg, et al., 2010). Here, we test if high-speed slice-accelerated multiband EPI sequence can help to leverage the dynamic content of resting-state BOLD signal for the purpose of inferring effective connectivity network using SEM.

## Materials and Methods

The fMRI scans were performed on eight healthy and right-handed subjects, according to the guidelines set forth by the institutional review board. They were scanned on a Siemens 3T Trio/Tim system using a multiband EPI sequence (Feinberg, et al., 2010). Acquisition parameters were field of view (FOV) = 224 mm, matrix = 64 × 64, echo time (TE) = 25 ms, flip angle = 60° and forty-four axial slices (3 mm thick without gap). Four subjects were scanned with repetition time (TR) = 1 sec and slice acceleration factor = 2, and four subjects were scanned with TR = 549 ms and slice acceleration factor = 4. For both groups, the resting-state fMRI scan took about 4 minutes. Subjects were instructed to close their eyes but stay awake.

For each fMRI data set, images were realigned to correct head motion. Thereafter, the data were regressed with the average signals from CSF and white matter to remove the physiological noises. Aided by the anatomical landmarks, we selected the areas that were closest to the coordinates cited in other studies to localize the left and right primary motor areas (L/R M1), the posterior cingulate cortex (PCC) and inferior parietal lobules (IPL) as ROIs. The time series of the voxels from each subject's identified ROIs were first averaged, and then normalized into percentage scale by subtracting and dividing by the mean. To explain the whole analysis procedure, we will start with one model as an example. In this model, two pathways start from the unknown region to LM1 and RM1 areas (Figure 1, Right Model) for the detection of sensory-motor network. The normalized time courses from LM1 and RM1 areas were treated as two observed variables. The signal from each voxel within the brain except the marked LM1 and RM1 areas was regarded as the third observation in the current structural equation model, and was evaluated with the SEM statistics. The time series of each voxel as the third variable has a time shift $\Delta t$ iterated between 0, -1TR and -2TR. Another similar model in this study is: two pathways start from the unknown region to PCC and IPL areas (Figure 1, Left Model) for the detection of sensory-motor network. All these two models consists of two connections start from the unknown region to the two predefined ROIs.

For the assessment of the overall model fit, the goodness of fit index (GFI) and the adjusted goodness of fit index (AGFI) are the most commonly used fit indices in SEM analysis. For our purpose, AGFI is a more appropriate index of fit than GFI, in that AGFI accounts for the number of degrees of freedom in the model (see Equations (1) and (2); Byrne, 1994; Gerbing and Anderson, 1993; Hu and Bentler, 1999). These indices are derived by

$$\text{GFI} = 1 - \text{tr}[(S^{-1}S-I)^2]/\text{tr}[(S^{-1}S)^2] \tag{1}$$

where tr indicates the trace operation, S is the covariance matrix of signals, and S is the estimated S;

$$\text{AGFI} = 1 - [p(p+1)/2df](1-\text{GFI}) \tag{2}$$

where p the number of observations, and df the degree of freedom in the model.

The SEM software Lisrel 8 (Jöreskog and Sorbom, Scientific Software International Inc., Chicago, IL) was used to estimate the statistical significance of model fitting with the experimental data at each voxel. The AGFI value found in the SEM analysis was marked on the image voxel when two statistical significance thresholds in the model were reached at that point. First, a threshold of 0.90 was chosen for the overall model significance, represented by the AGFI. Second, the significance of each path, or the t-value of each estimated path coefficient in the model was thresholded at 1.96, which corresponds to a probability level of 0.05 for the given degrees of freedom. The algorithm was implemented in Matlab (MathWorks, Natick, MA). Thereafter the model fit indices were compared between connectivity maps obtained from different temporal resolution scans.

## Results and Discussion

The AGFI map from sensory-motor network were determined and displayed in Figure 1 (right). The results of fitting the sensory-motor network model to all eight subjects' data demonstrated significant AGFI areas located in supplementary motor area (SMA), which is consistent with the well-known anatomical model of the two pathways. Table 1 lists the results of two models fitting in eight subjects. Meanwhile, the default mode network model fit best in the medial prefrontal cortex (MPFC) for all eight subjects.

The connectivity maps were found to be reproducible across subjects at the higher temporal resolution (TR= 549 ms), but not at the lower temporal resolution (TR=1s). With the higher temporal resolution, these results are consistent with the neuroanatomical evidence and existing results from non-directional functional connectivity data. Parts of predefined ROIs shown in the connectivity map can be interpreted as direct interactions between the two ROIs, which is also often found in the previous studies of functional connectivity. The largest standardized residual (the latent variable) obtained from SEM fitting at lower temporal resolution is much larger than that obtained at higher temporal resolution (Table 1), which suggests high-temporal-resolution data enabled by multiband fMRI can provide more robust information about the dynamic characteristics of connectivity networks.

As described in the introduction, standard methods of analyzing BOLD fMRI are not suitable to make definitive conclusions regarding the causal connectivity based on the observed temporal differences in the fMRI activations as the hemodynamic response is too slow, while the cortical inter-connection exhibited by neurophysiological oscillation on the millisecond scale can be studied by EEG or MEG (Darvas et al., 2004; David et al., 2006). The SEM approach is also applicable in analyzing EEG or MEG data in this regard (Babiloni et al., 2003; Astolfi et al., 2005). However, in general, neither EEG nor MEG has sufficiently high spatial resolution to reveal detailed cortical pathways. Therefore, we propose based on our current results that fMRI can serve as a complementary method for studying neural functions and neural interactions at a high spatial resolution using voxel-by-voxel SEM, especially with multiband fMRI data.

In our present study, the SEM enabled the use of time series from various groups of voxels to predict the interconnection of pathways that correspond to the sensory-motor network and default-mode network. Simple t-tests or cross correlation are insufficient to detect areas with a complex connectivity pattern. However, because the BOLD signal in particular areas, which have connections to any known areas, must follow a certain modulation, we can circumnavigate this problem by estimating different structural equation models. In this way, the present exploratory SEM application represents a new mapping method for revealing the causal influence of "latent" or "hidden" areas, and further provides information about the connectivity at each voxel.

The present approach of SEM analysis on the multiband fMRI data allows us to search the possible effective connections at each brain region directly from high temporal resolution signals. When several areas are involved in a network, which is often the case, the connections can become too complicated to be ascertained simply via correlation analysis. In contrast, SEM produces directional maps and estimations that circumvent this difficulty.

Multiband fMRI enables data acquisition at high temporal resolution and allows more applications of SEM in inferring effective connectivity.

**References**

Astolfi, L., Cincotti, F., Babiloni, C., Carducci, F., Basilisco, A., Rossini, P.M., Salinari, S., Mattia, D., Cerutti, S., Dayan, D.B., Ding, L., Ni, Y., He, B., Babiloni, F., 2005. Estimation of the cortical connectivity by high-resolution EEG and structural equation modeling: simulations and application to finger tapping data. IEEE Trans. Biomed. Eng. 52, 757–768.

Büchel, C., Friston, K.J., 1997. Modulation of connectivity in visual pathways by attention: cortical interactions evaluated with structural equation modelling and fMRI. Cereb. Cortex 7, 768–778.

Babiloni, F., Cincotti, F., Basilisco, A., Maso, E., Bufano, M., Babiloni, C., Carducci, F., Rossini, P., Cerutti, S., Rubin, D.B.D., 2003. Frontoparietal cortical networks revealed by structural equation modeling and high resolution EEG during a short term memory task. Proceedings of First International IEEE EMBS Conference on Neural Engineering, 20, pp. 79–82.

Byrne, B.M., 1994. Structural equation modeling and EQS and EQS windows: basic concepts, applications, and programming. Sage Publications, Thousand Oaks, CA.

Darvas, F., Pantazis, D., Kucukaltun-Yildirim, E., Leahy, R.M., 2004. Mapping human brain function with MEG and EEG: methods and validation. NeuroImage 23, S289–S299.

David, O., Kiebel, S.J., Harrison, L.M., Mattout, J., Kilner, J.M., Friston, K.J., 2006. Dynamic causal modeling of evoked responses in EEG and MEG. NeuroImage 30, 1255–1272.

Feinberg, D.A., Moeller, S., Smith, S.M., Auerbach, E., Ramanna, E., Glasser, S., Miller, K.L., Ugurbil, K., Yacoub, E., 2010. Multiplexed echo planar imaging for sub-second whole brain fMRI and fast diffusion imaging. PLOS One 5(12), e15710.

Gerbing, D., Anderson, J., 1993. Monte Carlo evaluations of goodness-of-fit indices. In: Bollen, K.A., Long, J.S. (Eds.), Testing Structural Equation Models. Sage Publications, Newbury Park, CA, pp. 40–65.

Goncalves, M.S., Hall, D.A., 2003. Connectivity analysis with structural equation modelling: an example of the effects of voxel selection. NeuroImage 20, 1455–1467.

Hu, L., Bentler, P.M., 1999. Cutoff criteria for fit indexes in covariance structure analysis: conventional criteria versus new alternatives. Struct. Equ. Modeling 6, 1–55.

McIntosh, A.R., Gonzalez-Lima, F., 1994. Structural equation modeling and its application to network analysis in functional brain imaging. Hum. Brain Mapp. 2, 2–22.

Penny, W.D., Stephan, K.E., Mechelli, A., Friston, K.J., 2004. Modelling functional integration: a comparison of structural equation and dynamic causal models. NeuroImage 23, S264–S274.

Smith, J.F., Chen, K., Johnson, S., Morrone-Strupinsky, J., Reiman, E.M., Eric, M., Nelson, A., Moeller, J.R., Alexander, G.E., 2006. Network analysis of single-subject fMRI during a finger opposition task. NeuroImage 32, 325–332.

Zhuang, J., LaConte, S., Peltier, S., Zhang, K., Hu, X., 2005. Connectivity exploration with structural equation modeling: an fMRI study of bimanual motor coordination. NeuroImage 25, 462–470.

Zhuang, J., Peltier, S., He. S., LaConte, S., Hu, X., 2008. Mapping the connectivity with structural equation modeling in an fMRI study of shape-from-motion task. NeuroImage 42, 799–806.

| TR (ms) | 549 | 1000 |
|---|---|---|
| **Sensory-motor Network** | 0.0154 (0.0083) | 0.0346 (0.0104) |
| **Default-mode Network** | 0.0181 (0.0096) | 0.0579 (0.0133) |

**Table 1.** Largest standardized residual averaged (and its standard deviation) from SEM fittings at each network and each scan.

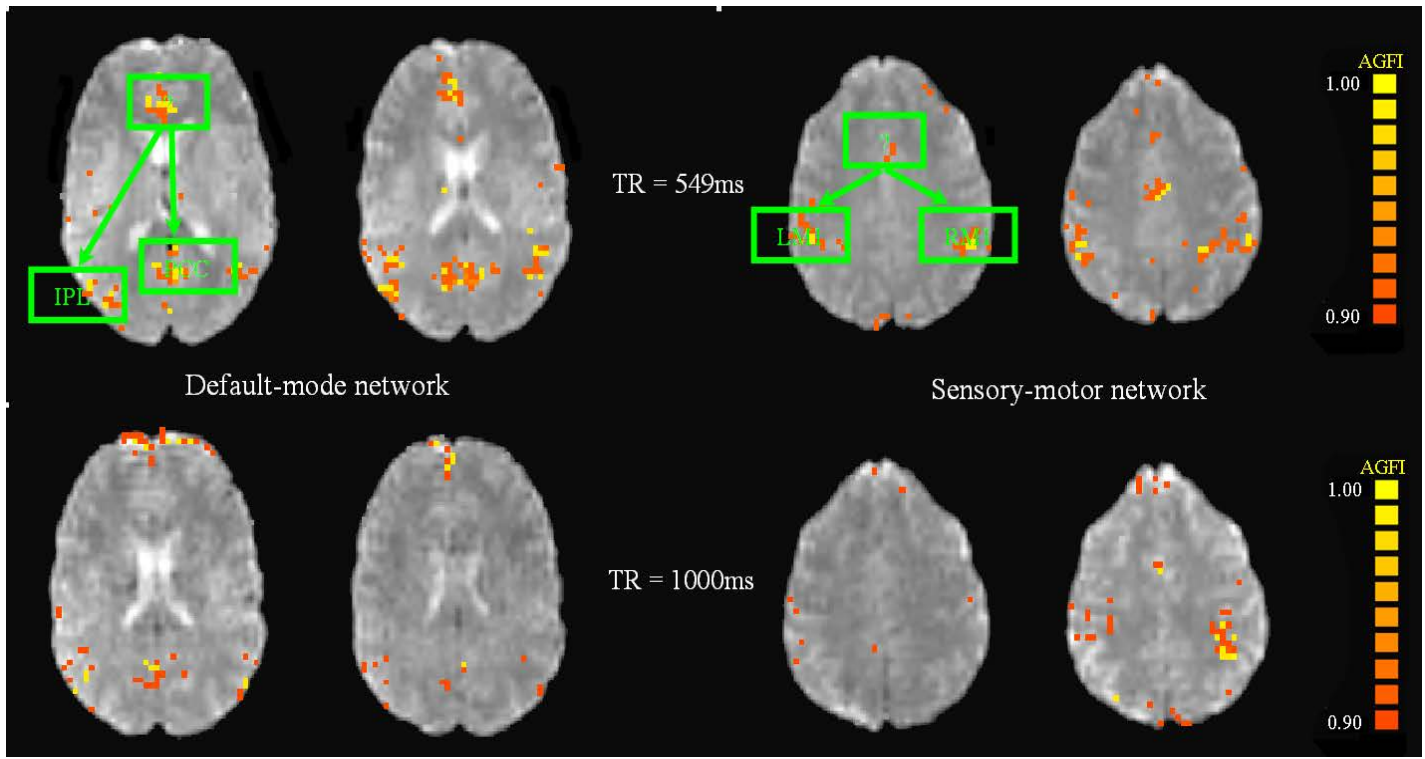

**Figure 1.** The structural equation models (in green) and resulting connectivity maps on two typical data from multiband fMRI.